\documentclass[12pt]{article}
\usepackage{epsfig}
\begin{document}
\begin{center}

{\bf ANALYTICAL AND NUMERICAL VERIFICATION OF THE NERNST THEOREM
FOR METALS }

\vspace{1cm}
 Johan S. H{\o}ye\footnote{E-mail:
johan.hoye@phys.ntnu.no}

\bigskip

Department of Physics, Norwegian University of Science and
Technology, N-7491 Trondheim, Norway

\bigskip

Iver Brevik\footnote{E-mail: iver.h.brevik@ntnu.no}

\bigskip

Department of Energy and Process Engineering, Norwegian University
of Science and Technology, N-7491 Trondheim, Norway

\bigskip

Simen A. Ellingsen\footnote{E-mail: simen.ellingsen@kcl.ac.uk}

\bigskip

Department of War Studies, King's College London, Strand, London
WC2R 2LS, UK

\bigskip

Jan B. Aarseth\footnote{E-mail: jan.b.aarseth@mtf.ntnu.no}

\bigskip

Department of Structural Engineering, Norwegian University of
Science and Technology, N-7491 Trondheim, Norway

\bigskip

\today
\end{center}

\bigskip

\begin{abstract}

In view of the current discussion on the subject, an effort is
made to show very accurately both analytically and numerically how
the Drude dispersion model gives consistent results for the
Casimir free energy at low temperatures. Specifically, for the free
energy near $T=0$ we find
the leading term proportional to  $T^2$ and
the next-to-leading
term proportional to $T^{5/2}$.
These  terms give rise to zero
Casimir entropy as $T \rightarrow 0$,   and is  thus in accordance with
Nernst's theorem.

\end{abstract}


\section{Introduction}

In recent years there has been a lively discussion about the
thermodynamic consistency of the expression for the Casimir
pressure at finite temperature $T$. The problem gets accentuated
at low values of $T$, where one has to satisfy the Nernst theorem
saying that $S=-\partial F/\partial T$ goes to zero as
$T\rightarrow 0$. (Here $S$ is the entropy and $F$ the free
energy, both referring to unit plate area.) What we shall consider
in the following is the standard Casimir configuration, implying
two semi-infinite homogeneous metallic media separated by a vacuum
gap of width $a$. We take the two media to be identical, and
assume that they are nonmagnetic with a frequency dependent
 relative permittivity $\varepsilon(\omega)$.  Spatial dispersion
is neglected. The two surfaces lying at $z=0$ and $z=a$ are
assumed to be perfectly plane, and to be of infinite extent.

A central ingredient in the discussion of the thermodynamic
consistency of  calculated results for the Casimir attractive
force between real materials is the form of dispersion relation
used  as input in the   conventional Lifshitz formula. A very
 useful dispersion relation -  the one that in our opinion is
  by far the most preferable one amongst simple dispersion relations
 for real systems at arbitrary frequencies  - is the
Drude expression
\begin{equation}
\varepsilon(i\zeta)=1+\frac{\omega_p^2}{\zeta(\zeta+\nu)}.
\label{1}
\end{equation}
Here $\omega=i\zeta$, $\omega_p$ is the plasma frequency, and
$\nu$ is the relaxation frequency (we use the same notation as in
Ref.~\cite{hoye03}). The plasma wavelength is $\lambda_p=2\pi
c/\omega_p$. For gold, the substance that we shall focus on in the
following, we use
\begin{equation}
\omega_p=9.03\, {\rm eV},\quad \nu=34.5\,{\rm meV},\quad
\lambda_p=137.4\,{\rm nm}. \label{2}
\end{equation}
In Ref.~\cite{hoye03}, we employed the values $\omega_p=9.0$ eV,
$\nu=35$ meV, what amounts roughly to a difference on the 1\%
level. The exact determination of Drude parameters is a
non-trivial matter as discussed in \cite{pirozhenko06}.
 Using a slightly different set of Drude parameters will shift our numerical results slightly,
  but does not alter any of our  conclusions.

When comparing with experimental values it turns out that the
Drude relation fits optical data very accurately for $\zeta <
2\times 10^{15}$ rad/s \cite{lambrecht00,lambrecht00a}. In this
connection we should bear in mind the following fact (cf. also the
discussion in Ref.~\cite{brevik06}): There exist no measurements
of the permittivity at very low frequencies. What is available, is
a series of measurements  of the imaginary part
$\varepsilon''(\omega)$ of the complex permittivity
$\varepsilon(\omega)=\varepsilon'(\omega)+i\varepsilon''(\omega)$.
The Kramers-Kronig relations then permit us to calculate the real
part $\varepsilon'(\omega)$, and thus the complete
$\varepsilon(\omega)$ is known.  Permittivity data kindly supplied
by Astrid  Lambrecht cover a very large frequency region, from
$1.5\times 10^{11}$ rad/s to $1.5\times 10^{18}$ rad/s. From these
data, the relaxation frequency $\nu$ can be derived. As mentioned,
from comparison with experimental data  it turns out that $\nu$
can be taken to be constant to a good accuracy, up to about $\zeta
= 2\times 10^{15}$ rad/s.
 For low frequencies  the Drude relation yields the proper extrapolation
down to $\zeta =0$.

A word of caution is  called for, as regards the circumstance that
permittivity measurements are done at {\it room temperature} in
practice. The frequency $\nu$ is in principle temperature
dependent, and we do not know the value of $\nu(T=0)$ very
accurately. It might seem natural here to invoke the
Bloch-Gr\"{u}neisen formula for the temperature dependence of the
electrical resistivity (cf.~\cite{handbook} or also the discussion
in Ref.~\cite{hoye03}).  From this, a form for $\nu=\nu(T)$ can in
principle be found.  According to the formula,
 the value of $\nu$ should go to {\it zero} as
$T\rightarrow 0$.  However, in practice this is not true. There
are always impurities present, which make the value of $\nu$
finite at $T=0$ \cite{khoshenevisan79}. The Bloch-Gr\"{u}neisen
formula, thus, is not followed in this limit. Mathematically, the
important point is that
\begin{equation}
\zeta^2[\varepsilon(i\zeta)-1] \rightarrow 0 \quad {\rm as} \quad
\zeta\rightarrow 0. \label{3}
\end{equation}
This relation ensures that the zero-frequency transverse electric
(TE) mode does not contribute to the Casimir force at all, as
discussed in detail in Ref.~\cite{hoye03}.
Strictly, the Drude parameters of Eq.\ (\ref{2}) are valid at room-temperature, and $\nu$ will take significantly smaller values for low temperatures. This affects our numerical results quantitatively, but not qualitatively; as long as $\nu$ is nonzero, the TE part of the free energy vanishes at zero frequency which is the central point.

The recent series of works on the Casimir effect by the present
group of authors
\cite{hoye03,brevik04,brevik05,hoye06,brevik06,brevik06a,ellingsen07,ellingsen07a}
- built upon the Lifshitz formula and the measured values of
$\varepsilon(\omega)$ in combination with the Drude relation -
have nowhere been found to run into conflict with basic
thermodynamic principles. And there are other papers in agreement
with ours: for instance, Jancovici and \v{S}amaj
\cite{jancovici05}, and Buenzli and Martin \cite{buenzli05}
considered the Casimir force between two plates in the
high-temperature limit. They found that the linear dependence in
$T$ for the Casimir force to be reduced by a factor of two from
the behavior of of an {\it ideal} metal, this being a signal of
the vanishing influence from the zero-frequency TE mode. (The
first observation of the vanishing influence from this particular
mode was made by Bostr{\"o}m and Sernelius \cite{bostrom00}.)
Further support is found in the paper of Sernelius
\cite{sernelius06}, who calculates the Casimir force taking
spatial dispersion into account as well. It is found that at high
temperatures and/or at large separations the force is reduced by
the same factor of two compared with the ideal-metal result.

There is no universal agreement on these issues, however. In a
series of recent papers - cf., for instance,
Refs.~\cite{bezerra04,decca05,bezerra06,mostepanenko06,mostepanenko07}
- it is  argued that the Drude dispersion relation runs  into
trouble explaining recent experiments, and moreover comes into
conflict with the Nernst theorem. These authors favor, instead of
the Drude relation, the plasma relation
\begin{equation}
\varepsilon(i\zeta)=1+\frac{\omega_p^2}{\zeta^2}, \label{4}
\end{equation}
which corresponds to setting $\nu=0$ in Eq.~(\ref{1}). (It should
be noted that the expression (\ref{4}) does {\it not} satisfy the
condition (\ref{3}).)

An argument of the latter references is that omission of a zero
frequency TE-mode would add a term linear in $T$ to the free
energy. This would  violate the Nernst theorem as it would give a
non-zero contribution to the entropy at $T=0$. However, this
argument is based upon use of the IM (idealized metal) model where
$\varepsilon=\infty$ for all $\zeta$, or use of the plasma model
(\ref{4}) where no relaxation is present. With $\nu>0$ the term in
question is still linear away from $T=0$, but the precise behavior
as $T\rightarrow0$ has been less obvious. As argued in
Ref.~\cite{hoye03} the straight line should bend to become
horizontal at $T=0$. This was not verified in utmost detail,
however; the numerical results of Ref.~\cite{hoye03} did not go
sufficiently close to $T=0$ to show the behavior very distinctly,
and the previous discussion  and disagreement  about violation of
the Nernst theorem has accordingly continued. The main purpose of
the present work is to investigate the issue more closely: we will
show in detail, both analytically and numerically, how the
Casimir energy behaves close to $T=0$ and by that show how it is
consistent with the Nernst theorem.

We shall not go into a study of experimental aspects in this
paper. Rather, the objections referred to above makes it mandatory
to reconsider the thermo{\-}dynamics associated with the Drude
relation anew, which brings us to the  central theme of this
paper. We will aim at showing, via a combination of analytical and
numerical methods, how the basic theory sketched above
(essentially the Drude theory) satisfies the Nernst theorem to a
very high accuracy. We consider this point to be important; a
simple physical model  of course cannot be permitted to run into
conflict with  thermodynamics.

In the next section we show analytically, by using the
Euler-Maclaurin formula, that the dominant contribution to the
free Casimir energy $F$ near $T=0$ is proportional to $T^2$. We
evaluate both this term and the leading correction term, which is
proportional to   $T^{5/2}$. This implies that the entropy
$S=-\partial F/\partial T$ tends to zero as $T\rightarrow 0$, in
accordance with the Nernst theorem. In Sect. 3 we calculate $F$
numerically, and find agreement with the previous analytical
result to a very high degree of accuracy. The results are
illustrated in various figures. Thus we can conclude that the
Drude ansatz   does not run into conflict with thermodynamics
at all.

Readers interested in recent reviews on the Casimir effect may
consult Milton's book\footnote{It may  be mentioned for
completeness that this book from 2001 was written from the
standpoint of the "classical" IM model. } \cite{milton04}, and
several review articles
\cite{bordag01,milton04a,nesterenko04,lamoreaux05}. A great deal
of recent information can also be found in the special issues of
J. Phys. A: Math. Gen. \cite{jpa06} and of New J. Phys.
\cite{njp06}.

As mentioned above, we shall not be concerned with comparison
between theory and experiment in the present paper. We mention,
though, the recent experiment of Obrecht {\it et al.}
\cite{obrecht07}, which seems to report the first accurate
measurement of thermal Casimir effects. The experiment is
important, but it lies outside the scope of the present
investigation since it deals with non-uniformly heated systems.

Finally, we mention the special variant of the thermal Casimir
problem consisting in studying, instead of a metal, a {\it
semiconductor} endowed with a  small but finite conductivity at
zero frequency \cite{geyer05,klimchitskaya06}.  According to the
authors of these references this  situation implies  an
interchange of the roles of the TE and TM (transverse magnetic)
modes, as compared with the case  of a metal. Namely, within an
idealized approach, they find  that the TM reflection coefficient
gets a discontinuity at $\zeta=0$, implying in turn an apparent
conflict with the Nernst theorem. The problem is interesting, and
we hope to return to it in a later paper.

\section{Analytical approach: Casimir free energy near $T=0$ for real metals}
\label{sec2}
As mentioned, we use the same notation as in Ref. \cite{hoye03}.
We will evaluate the leading $T$-dependence of the Casimir free
energy near $T=0$ for metals using the Drude relation (\ref{1}),
\begin{equation}
\varepsilon(i\zeta)-1=\frac{\omega_p^2}{\zeta(\zeta+\nu)}
\approx \frac{D}{\zeta}, \quad {\zeta \ll \nu},
\label{5}
\end{equation}
with $D=\omega_p^2/\nu$. The free energy is given by expression
(3.4a) in \cite{hoye03} as
\begin{equation}
\beta
F=\frac{1}{2\pi}{\sum\limits_{m=0}^\infty}^{\prime}\int\limits_{\zeta/c}^\infty
[\ln(1-\lambda^\mathrm{TM})+\ln(1-\lambda^\mathrm{TE})]qdq \label{6}
\end{equation}
where $\beta = 1/kT$ and
\begin{eqnarray*}
  \lambda^\mathrm{TM}&=&Ae^{-2qa},\nonumber\\
  \lambda^\mathrm{TE}&=&Be^{-2qa}.
\end{eqnarray*}
The prime on the summation sign means that the case $m=0$ is taken
with half weight. The coefficients $A$ and $B$ are the squared
Fresnel reflection coefficients for the two media and are given by
\begin{equation} \label{AB}
  A=\left(\frac{s-\epsilon p}{s+\epsilon p}\right)^2,\quad
  B=\left(\frac{s-p}{s+p}\right)^2,
\end{equation}
with
\[s=\sqrt{\varepsilon-1+p^2},\quad p=\frac{qc}{\zeta}. \]
  Here $a$ is the plate separation, $\varepsilon$  the relative permittivity,
   $c$  the velocity of light in vacuum, and $\zeta$ is the Matsubara
frequency given by

\begin{equation}
  \zeta=\frac{2\pi k}{\hbar}mT. \label{7}
\end{equation}
 Note that the quantities of $A,B,s,p$ and $\zeta$ all depend upon the summation variable $m$.  (Units $c=\hbar=k=1$ are not used.) The term of interest is the TE
mode, since this is the term  that  gives rise to the controversy about
the Nernst theorem.

 With the small $\zeta$ dependence of Eq.~(\ref{5}) the $B$ has a scaling
form such that it can be expressed in terms of one variable. So by introducing a new
variable $x$ to replace $q$ the $\zeta$-dependence can be removed
fully
\begin{equation}
x^2=\frac{q^2 c^2}{(\varepsilon-1)\zeta^2}\approx \frac{q^2
c^2}{D\zeta}, \quad \zeta \ll \nu. \label{8}
\end{equation}
With this we have
\begin{equation}
B=\left(\frac{\sqrt{1+x^2}-x}{\sqrt{1+x^2}+x}\right)^2
=(\sqrt{1+x^2}-x)^4, \label{9}
\end{equation}
and the TE free energy expression can be written as

\begin{equation}
\beta F^{TE} =C{\sum\limits_{m=0}^\infty}^\prime g(m), \label{10}
\end{equation}
with
\begin{equation}
g(m)=m \int_{x_0}^\infty x\ln{(1-Be^{-y})}\,dx, \label{11}
\end{equation}
where
\begin{equation}
C=\frac{D}{c^2 \hbar\beta}=\frac{\omega_p^2}{c^2\hbar \nu\beta},
\quad y=2qa=2\frac{a}{c}\sqrt{D\zeta}\,x, \quad
x_0=\sqrt\frac{\zeta}{D}. \label{12}
\end{equation}

The small $T$-behavior can now be obtained by use of the
Euler-Maclaurin summation formula
\begin{equation}
{\sum_{n=0}^\infty}' g(n)=\int\limits_0^\infty g(u)\,du
-\frac{1}{12}g'(0)+\frac{1}{720}g'''(0)-... \label{13}
\end{equation}
One easily sees that
\begin{equation}
g'(0)=\int_{0}^\infty x\ln{(1-B)}\,dx. \label{14}
\end{equation}
This integral can be performed analytically. First introduce a new
variable $x=\sinh t$ with $dx=\cosh t \,dt$
\begin{equation}
g'(0)=\int\limits_0^\infty \sinh t \,\cosh t
\,\ln(1-e^{-4t})\,dt=\frac{1}{4}\int\limits_0^\infty
(e^{2t}-e^{-2t})\ln(1-e^{-4t})\,dt. \label{15}
\end{equation}
Next we substitute $e^{-2t}=u$ for which $-2e^{-2t}\,dt=du$ and
use partial integration to obtain

\begin{eqnarray}
g'(0)&=&-\frac{1}{8}\int\limits_0^1
\left(1-\frac{1}{u^2}\right)\ln(1-u^2)\,du
\nonumber\\
&=&-\frac{1}{8}\left\{
\frac{1}{u}[(1+u)^2\ln(1+u)+(1-u)^2\ln(1-u)]-2u\right\}_0^1
\nonumber\\
&=&-\frac{1}{4}(2\ln2-1)\approx-0.09657. \label{16}
\end{eqnarray}

At $T=0$ the free energy is determined by the integral in
Eq.~(\ref{13}) (besides contribution from the TM-mode). For small
$T\rightarrow0$ the deviation from the $T=0$ value is thus
($\beta=1/kT$)
\begin{equation}
\Delta F^{TE}=\frac{C}{\beta}\left[-\frac{1}{12}g'(0)\right]
=\frac{C} {48\beta}(2\ln 2-1). \label{17}
\end{equation}
This result  was presented by Milton at the QFEXT03 Workshop, and
is given as Eq.~(22) in Ref.~\cite{brevik04}, or Eq.~(4.9) in
Ref.~\cite{brevik06a}.

It can be noted that $\Delta F^{TE}$ is independent of the plate
separation $a$ and can thus be valid only sufficiently close to
$T=0$ for a given $a$  such that $\Delta F^{TE}\ll F^{TE} \sim
\hbar c/a^3$. Evaluation of the next term as given by result
(\ref{29}) below verifies this. The leading term of $\Delta
F^{TE}$ dominates its next term only when the dimensionless
quantity $aC^{1/2}\ll 1\;$ or $a^2 kT \ll \hbar\nu
c^2/\omega_p^2$. A consequence of this is that for increasing $a$
the temperature interval where (\ref{17}) is valid decreases
rapidly, and $\Delta F^{TE}$ becomes more and more negligible
compared to $F^{TE}$ since then  $\Delta F^{TE} \sim
C/\beta=\omega_p^2 (kT)^2/(c^2 \hbar\nu) \ll c^2
\hbar\nu/(\omega_p^2 a^4)$. (Thus in the present case $\Delta
F^{TE}\ll F^{TE}$ provided $a>c\nu/\omega_p^2\approx 0.1$\,nm with
$a^2 kT \ll \hbar\nu c^2/\omega_p^2$ fulfilled.) In the limit
$a\rightarrow\infty$ the IM ($\varepsilon=\infty$), but now with
the linear term included, is recovered. But the latter does not
violate the Nernst theorem as long as $a$ is finite, and for
$a=\infty$ there is no Casimir free energy anyway\footnote{ It may
be noted that for $a/(\hbar c\beta)\ll 1$ an $a$-independent
contribution to the free energy was found also in the "classical"
Casimir theory for metals at low temperature. See Eq.~(3.38) in
Ref.~\cite{milton04} or Eq.~(3.24) in Ref.~\cite{hoye03}.}.

Inserting the value for $g'(0)$ and the values
$\hbar\omega_p=9.03$\,eV and $\hbar\nu=34.5$\,meV for Au  we find
with $C$ given by (\ref{12}) ($\hbar=1.0545\cdot10^{-34}$\,Js,
$k=1.381\cdot10^{-23}$\,J/K, $c=2.998\cdot10^8$\,m/s)
\begin{equation}
\Delta F^{TE}=C_1 T^2\quad \mbox{with}\quad
C_1\approx5.81\cdot10^{-13}\,\mbox{J/(m}^2{\rm K}^2). \label{18}
\end{equation}

It turns out that Eq.~(\ref{17}) holds only very close to $T=0$
(i.~e. $T\ll 0.01$\,K), but there will be a leading correction
that we can obtain with good accuracy. Expanding $g(m)$ in powers
of $m$ one notes that half integer powers will occur. Thus formula
(\ref{13}) is not quite valid as $g'''(0)$ and higher order
derivatives will diverge. However, this problem can be avoided
since the formula can be applied to summation starting at $m=p$
($p=1, 2, 3, \cdots$). Thus we have
\begin{eqnarray}
\sum_{n=0}^\infty g(n)&=&\sum_{n=0}^{p-1}
g(n)+\int\limits_p^\infty
g(u)\,du+\frac{1}{2}g(p)-\frac{1}{12}g'(p)+\frac{1}{720}g'''(p)+\cdots
\nonumber\\
&=&\int\limits_0^\infty
g(u)\,du+S-\frac{1}{12}g'(p)+\frac{1}{720}g'''(p)+\cdots
\label{19}
\end{eqnarray}
where
\begin{equation}
S=\sum_{n=0}^{p-1} g(n)+\frac{1}{2}g(p)-\int\limits_0^p g(u)\,du.
\label{20}
\end{equation}
For a power term  $g_\sigma(m)=m^\sigma$  (for small $m$) we have

\begin{eqnarray}
g'_\sigma(p)=\sigma p^{\sigma-1}, \quad
g'''_\sigma(p)=\sigma(\sigma-1)(\sigma-2)p^{\sigma-3},\quad
\int\limits_0^p g_\sigma(u) \,du=\frac{1}{1+\sigma}p^{\sigma+1}.
\label{21}
\end{eqnarray}
With the choice $p=1$ we get

\begin{equation}
S=S_\sigma(1)=\frac{1}{2}-\frac{1}{1+\sigma}. \label{22}
\end{equation}

One may note that  $S_1(1)=0$  as should be expected. The
power of key interest here will be $\sigma=3/2$ by which  $S_\sigma=1/10$ ,
and thus

\begin{equation}
\Delta S_\sigma(1) \equiv
S_\sigma(1)-\frac{1}{12}g'_\sigma(1)+\frac{1}{720}g'''_\sigma(1)\approx
-0.02552\approx-0.204\frac{1}{12}g'_\sigma(1). \label{23}
\end{equation}

When other terms are neglected the error can be estimated from the
next term in the series (provided the sum of higher order terms
converge)

\begin{equation}
\frac{1}{30240}g^{(V)}_\sigma(1)=\frac{-1}{30240}\,\frac{3}{2}\left(\frac{1}{2}\right)
\left(\frac{-1}{2}\right)\left(\frac{-3}{2}\right)\left(\frac{-5}{2}\right)\approx4.65\cdot10^{-5},
\label{24}
\end{equation}

which is only about 0.2\% of the value (\ref{23}). For increasing
values of $p$ the error goes further down rapidly. (Instead of
result (\ref{23}) $p=2$ yields $\approx-0.02549$.)  Thus
(\ref{23}) with $p=1$ is a good estimate for the sum of interest.

With $g(m)$ given by (\ref{11}) we can expand to leading order in
$m$ or $y ~(\propto T^{1/2})$. To this leading order the limit of
integration  $x_0 ~(\propto T^{1/2})$  can be put equal to zero since the integrand
vanishes for $x=0$. We find
\begin{eqnarray}
g(m)=mg'(0)+myI+\cdots \label{25}
\end{eqnarray}
Now from (\ref{7}) and (\ref{12})
\begin{equation}
y=2a\sqrt{2\pi C}xm^{1/2}, \label{26}
\end{equation}
with $C$ given by (\ref{12}). Thus the derivative becomes
\begin{equation}
g'(m)=g'(0)+3a\sqrt{2\pi C}\,Im^{1/2}+\cdots \label{27}
\end{equation}
where $g'(0)$ is given by (\ref{16}) and (with substitution
$x=\sinh t$ as in (\ref{15}))
\begin{equation}
I=\int\limits_0^\infty\frac{x^2
B}{1-B}\,dx=\frac{1}{8}\int\limits_0^\infty(e^{-t}-e^{-3t})\,dt=\frac{1}{12}.
\label{28}
\end{equation}
Now we  will  use Eq.~(\ref{20}) for the most simple case $m=p=1$
for which $S_\sigma(1)$ is given by (\ref{22}), and with exponent $\sigma$
equal to 1 and 3/2 for the two separate terms in (\ref{25}) the corresponding values of $S_\sigma(1)$ are 0
and 1/10 respectively. Thus with (\ref{23}) and (\ref{27}) we have
$12\,\Delta S_1 (1)=-g_1'(0)=-g'(0)$  and $12\,\Delta S_{3/2}(1)=-0.204\,g_{3/2}'(1)
=-0.204\cdot3a\sqrt{2\pi C} I$.
So with (\ref{17}) the change in free energy becomes
\[\Delta
F^{TE}=\frac{C}{\beta}[\Delta S_1 (1)+\Delta S_{3/2}(1)+\cdots]\]
\begin{equation}
= \frac{C}{\beta}\left[-\frac{1}{12}g'(0)\right]\left[1-0.204\cdot\frac{3a\sqrt{2\pi
C}}{-12g'(0)}+\cdots\right], \label{29}
\end{equation}
which gives the small $T$ correction to the result (\ref{17}).
Again inserting the previous values for $\omega_p$ and $\nu$ we
obtain with plate separation $a=1000$\,nm
\begin{equation}
\Delta F^{TE}=C_1 T^2[1-C_2 T^{1/2}+\cdots] \quad \rm{with} \quad
C_2\approx3.03\,\rm{K}^{-1/2}. \label{30}
\end{equation}

This result can only be valid for very small $T$ as it otherwise
becomes negative  already for $T$ slightly larger than 0.1\,K.
To avoid this one may instead write
\begin{equation}
\Delta F_{th}^{TE}=\frac{C_1 T^2}{1+C_2 T^{1/2}}. \label{31}
\end{equation}
as the theoretical or analytical result for $\Delta F^{TE}$ for
small $T$. This Pad\'{e} approximant form (which is equivalent to
Eq.~(\ref{30}) with respect to the first two terms), turns out to
be convenient for comparison with numerical evaluations. If the
corresponding numerical result for $\Delta F^{TE}$ is $\Delta
F_{num}^{TE}$ one can evaluate the ratio
\begin{equation}
R=\frac{\Delta F_{th}^{TE}-\Delta F_{num}^{TE}}{\Delta
F_{th}^{TE}}, \label{32}
\end{equation}
and consider the limit $T\rightarrow 0$ for which the limiting
value should be $R=0$ (cf. more details in Appendix B).

\section{Numerical calculation of free energy at low temperatures}

We have calculated the free energy $F^{TE}$ as a function of
temperature given by (\ref{6}) for two gold half-spaces separated
by a vacuum gap of width $a=1.0\,\mu$m. This would be a typical
experimental situation where the influence from the finite
temperature is large, about 15\% increase in the magnitude of the
Casimir free energy \cite{hoye03}, and a corresponding decrease of
the Casimir force according to our theory with no contribution
from the TE mode at zero frequency. The calculations use the
permittivity data for gold, received from Astrid Lambrecht as
mentioned. At $T=0$ the free energy is calculated numerically as a
double integral rather than a sum of integrals using a
two-dimensional  version of Simpson's method.

\begin{figure}
  \begin{center}
    a)\includegraphics[width=2.5in]{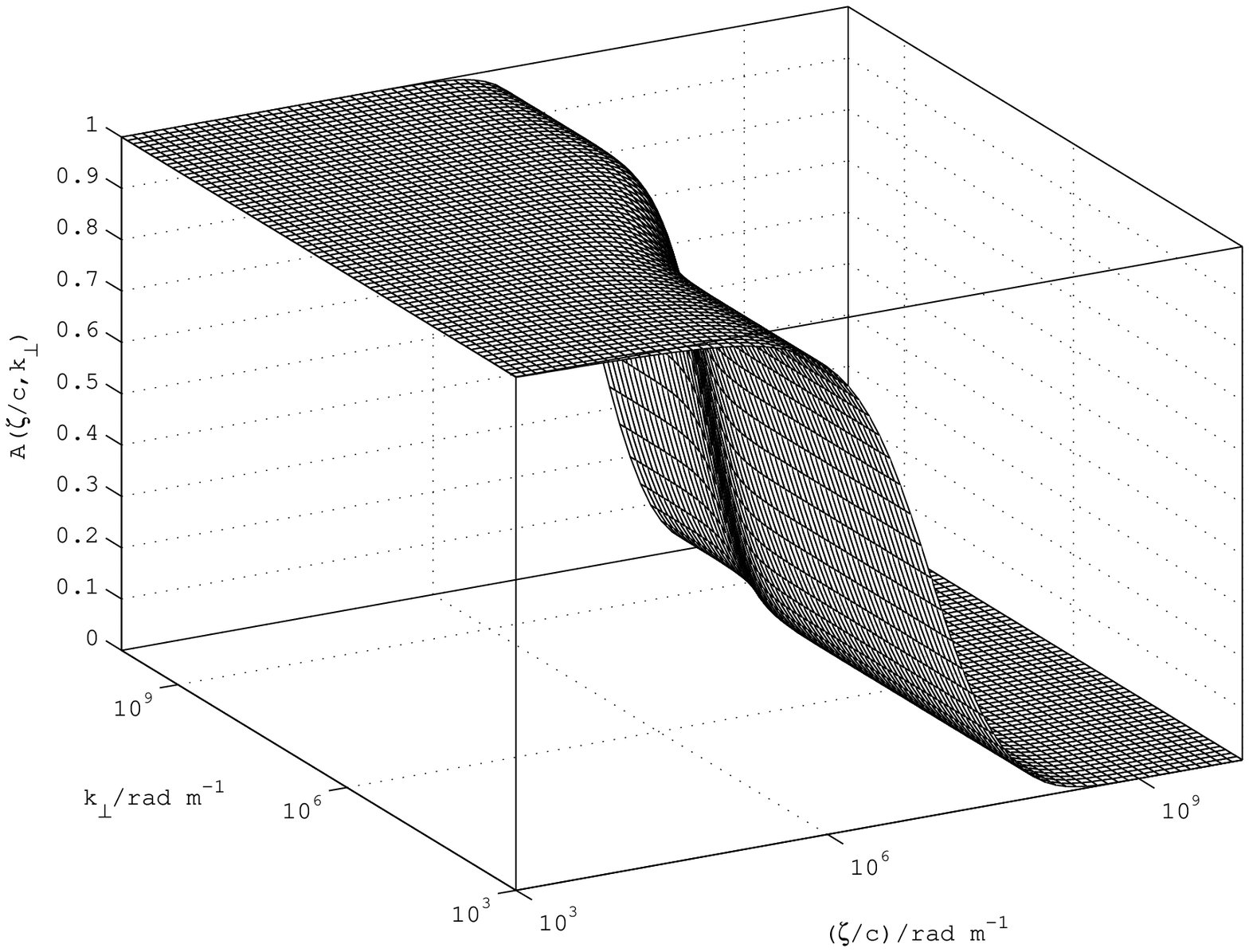}\\
    b)\includegraphics[width=2.5in]{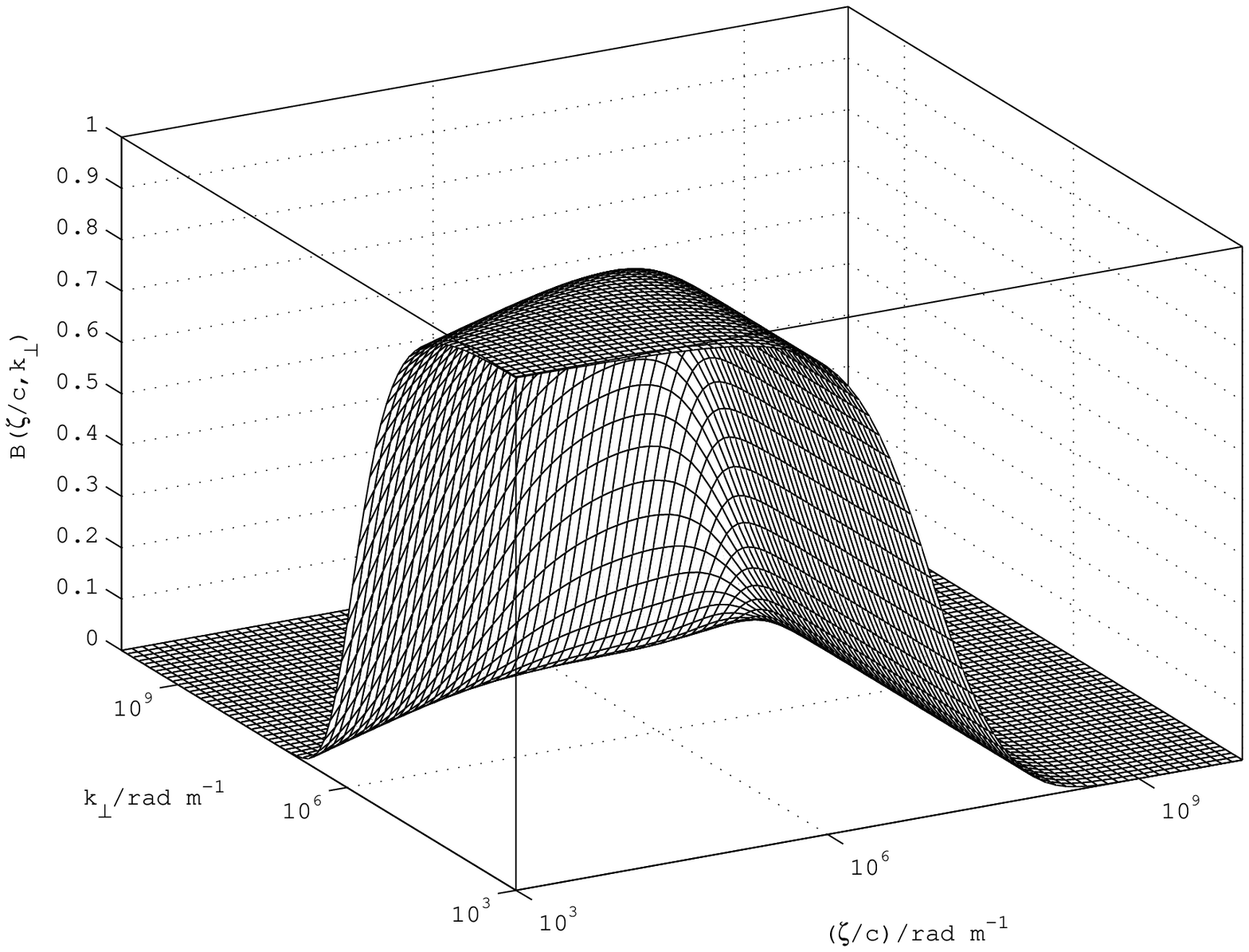}\\
    c)\includegraphics[width=2.5in]{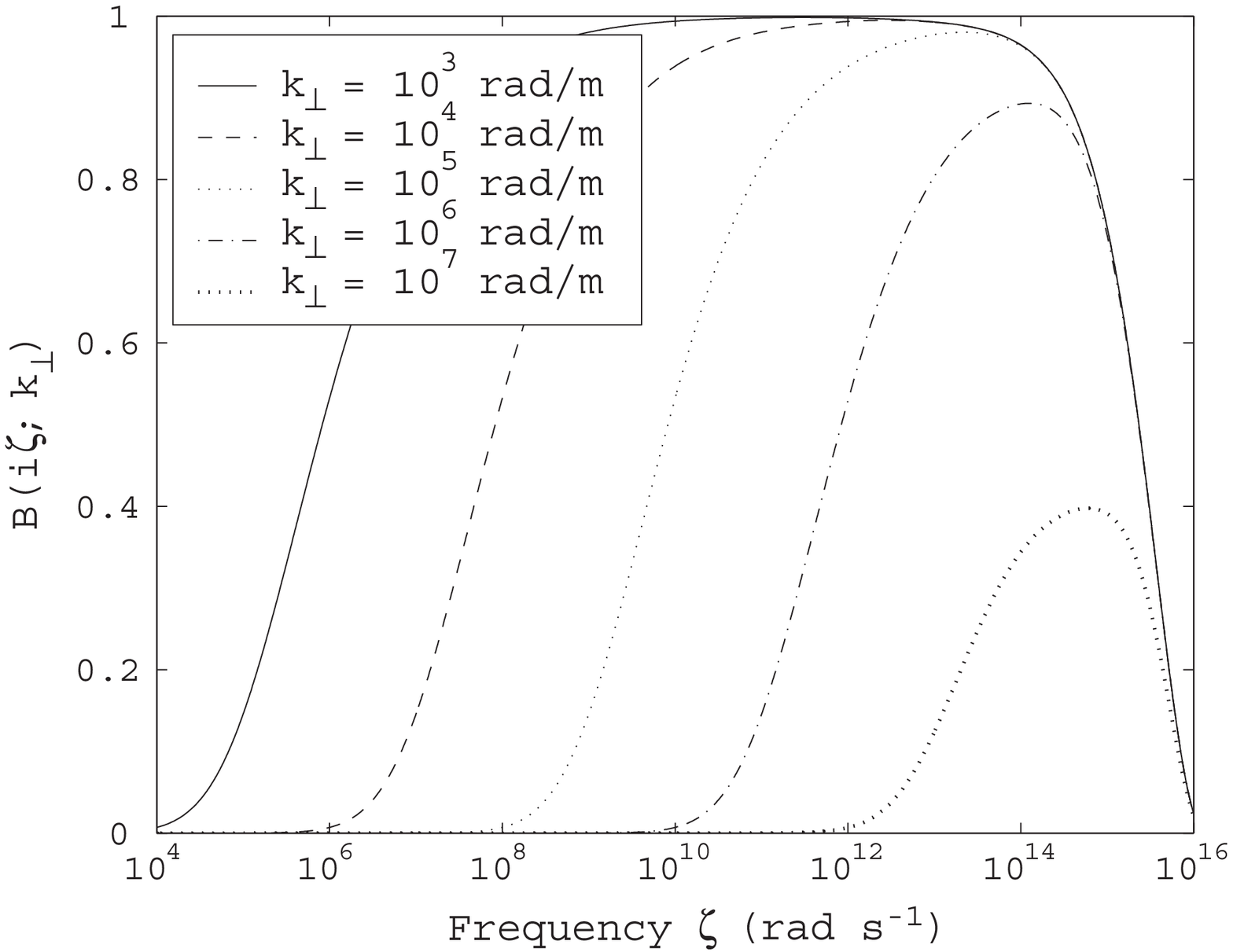}
    \caption{Squared reflection coefficients $A$  and $B$  of the metal interfaces as defined in (\ref{AB}) for the TM
    and TE modes,  as a function of $\zeta/c$ and the transverse momentum $k_\perp$. a) $A$ for the TM mode, b) $B$ for the TE mode, c) $B$ for $k_\perp$
   and $\zeta$ close to zero.}\label{fig_refl}
  \end{center}
\end{figure}

\begin{figure}
  \begin{center}
    \includegraphics[width=4.2in]{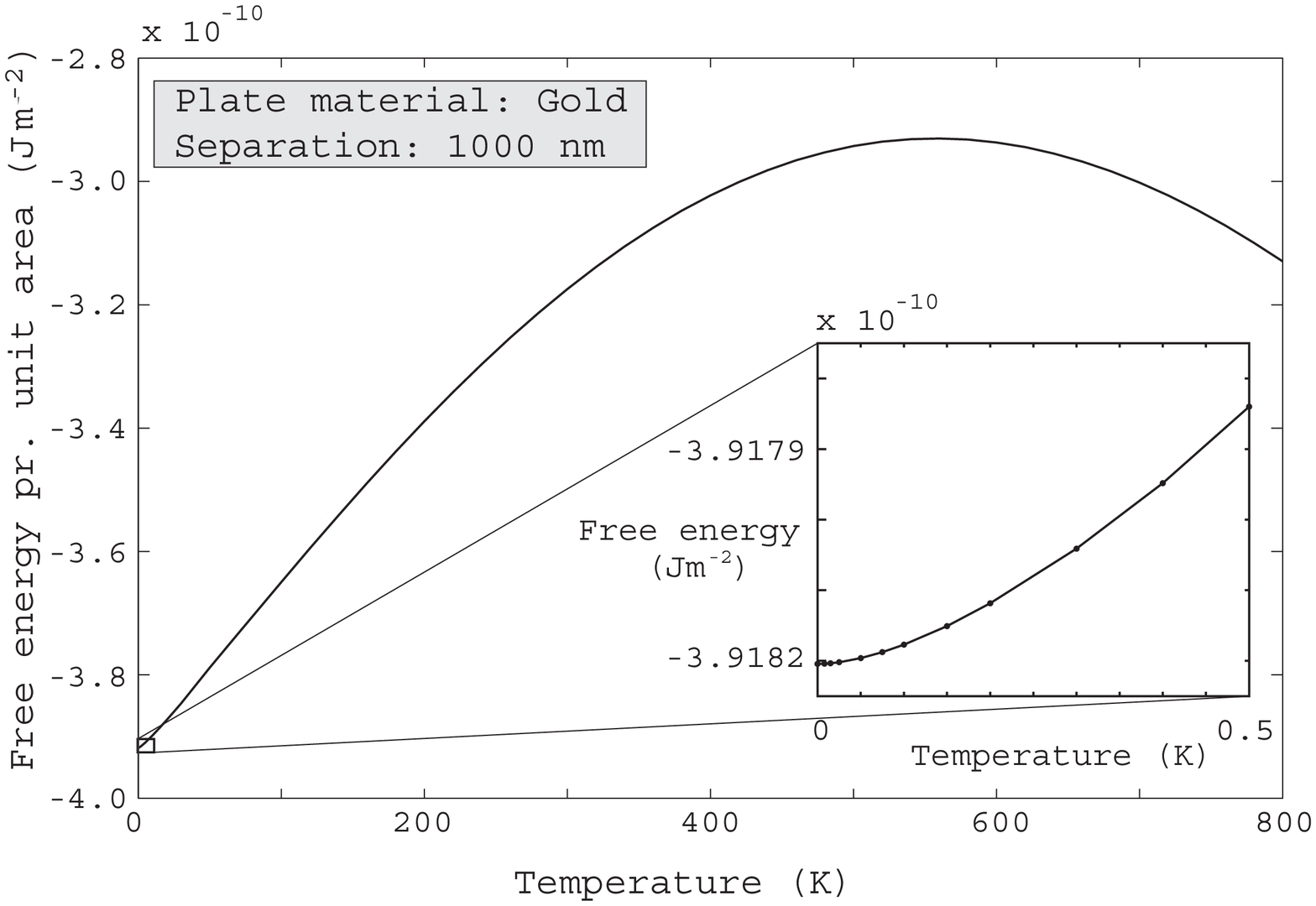}
    \caption{Numerical evaluation of the  free energy (\ref{6}) between two gold halfspaces as a function of temperature. The inset gives details for low $T$.}
    \label{fig_F(T)}
  \end{center}
\end{figure}

The vanishing of the zero frequency mode is intimately connected
with  the behavior of the reflection coefficient $B$ at vanishing
frequency. According to the Drude model (or any model satisfying
(\ref{3})) the TE mode reflection coefficient tends to zero as
$\zeta\to0$. To illuminate this point, we have plotted $A$ and $B$
as a function of  imaginary frequency and
transverse momentum, $k_\perp$, for an interface
between gold and vacuum in Fig.~\ref{fig_refl}. In   part c) of
this figure, we clearly see how $B$ vanishes smoothly when $\zeta
\rightarrow 0$ for $k_\perp \neq 0$ consistent with Maxwell's
equations of electrodynamics. However, the coefficient $A$ in
Fig.~\ref{fig_refl}a) for the TM mode equals 1 for all $k_\perp$
as $\zeta \rightarrow 0$, as is  also  the case for for all $\zeta$ for
an ideal metal. In the latter limit we also have $B\rightarrow 1$,
but for $\zeta =0$, $B$ remains zero.

For  ideal or non-ideal metals it is well known that the
temperature correction for the TM mode behaves as $T^4$. Thus it
does not add to the $T^2$ and $T^{5/2}$ terms that we find from
the TE  mode .

\begin{figure}
  \begin{center}
  \includegraphics[width=3.5in]{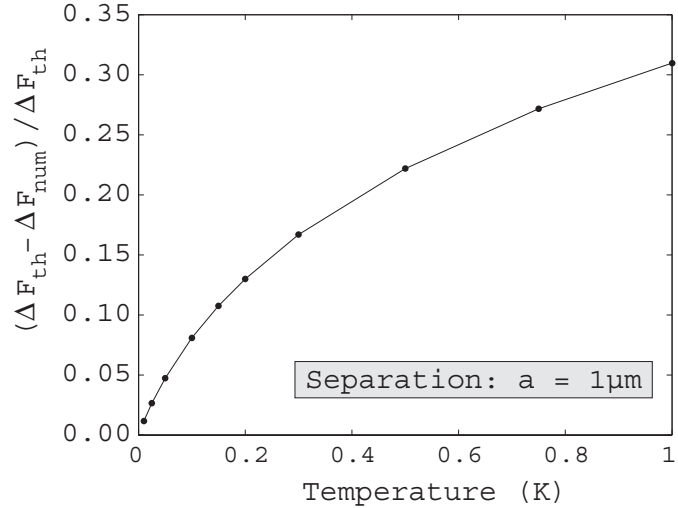}
   \caption{Plot of the ratio $R$ defined in Eq.~(\ref{32}). }\label{fig_numth}
  \end{center}
\end{figure}


By direct numerical integration and lengthy summations independent
of the analytic derivations made in Section \ref{sec2}, we obtain the free
energy numerically. Figure \ref{fig_F(T)} shows the free energy versus
temperature up to 800 K, while the  inset  shows details of the
parabolic shape close to $T=0$. First of all the figure shows the
controversial decrease of the  magnitude of the Casimir free energy and
thus also the related decrease of the Casimir  force up to a certain
temperature. Secondly, the  inset  shows that the tangent of the
curve is horizontal at $T=0$ as predicted. This implies that the
entropy due to the Casimir force is indeed zero at $T=0$. Thus the
Nernst theorem is not violated when using the realistic Drude
dispersion model. In contrast, it would be violated if a TE-term
were added for $\zeta=0$. This conclusion, as mentioned above,  is
clearly in contrast to that presented in  various  earlier works
\cite{bezerra04,decca05,bezerra06,mostepanenko06,mostepanenko07}.
The reader should note that the dependency of $\nu$ on temperature has been neglected in figure \ref{fig_F(T)}.

Now there are always some  uncertainties connected with numerical
calculations. Also the analytic derivation has some uncertainties
, e.g., concerning proper use of the Euler-Maclaurin formula, and
concerning convergence and neglect of of higher order terms. In
Fig.\ref{fig_numth}   we have therefore made a more accurate and much more
sensitive test of the behavior near $T=0$ comparing the analytic
result with the numerical one, by plotting the ratio $R$ defined
in Eq.~(\ref{32}).   We see that  $R$ when extrapolated approaches zero
linearly with a finite slope (when taking the curvature of the
plot into account). Thus, with high accuracy we find full
agreement concerning the $T^2$ and $T^{5/2}$ terms in the free
energy and their coefficients  $C_1$ and $C_2$.  As the number of terms to be summed
 numerically  increases rapidly when $T=0$ is approached, our evaluations were
terminated at $T=0.008~$K.  The extrapolated value $R=0$ for $T=0$
means that the coefficient $C_1$ is correct while the finite slope
means that $C_2$ is correct too within numerical uncertainties.
Also if a more dominant power were present, $R$ would diverge at
$T=0$. The finite slope of $R$ at $T=0$ means that the next term
in $\Delta F^{TE}$ is of higher order (see also Appendix B for
details).

\bigskip

{\bf Acknowledgment}

\bigskip

We thank Astrid Lambrecht for providing accurate dispersive data
for gold, and we thank Kimball A. Milton for valuable comments.

\renewcommand{\theequation}{\mbox{\Alph{section}\arabic{equation}}}
\appendix
\setcounter{equation}{0}

\section{Alternative derivation by expansion of $g(m)$}\label{app_A}

As a variant of the analytic approach, let us show how the
essential dependence of the free energy $F^{TE}$ on $T$ near $T=0$
  also can
be recovered  by making use of complex integration. We begin with
the TE expression
\begin{equation}
\beta F^{TE}= C{\sum_{m=0}^\infty}^\prime m \int_{x_0}^\infty x
\ln(1-Be^{-\alpha x})dx, \label{a1}
\end{equation}
where
\begin{equation}
 C=\frac{\omega_p^2}{\beta \hbar \nu c^2},
\quad \alpha=
2a\sqrt{2\pi Cm}.
 \label{a2}
\end{equation}
The essential step now is to expand the logarithm to the first
term,
\begin{equation}
\beta F^{TE}=-C\sum_{m=1}^\infty m\int_0^\infty x Be^{-\alpha
x}dx,\label{a3}
\end{equation}
($m=0$ does not contribute to the sum (\ref{a1})), where we have
also replaced the lower limit $x_0$ by zero. We next use the
formula \cite{whittaker62}
\begin{equation}
e^{-\alpha x}=\frac{1}{2\pi
i}\int_{c-i\infty}^{c+i\infty}ds(\alpha x)^{-s}\Gamma (s), \quad
c>0, \label{a4}
\end{equation}
 where $\Gamma(s)$ is the gamma function.  The summation over $m$ is easily done,
\begin{equation}
\sum_{m=1}^\infty m^{1-s/2} =\zeta(\frac{s}{2}-1).\label{a5}
\end{equation}
Here $\zeta$ is the Riemann zeta function, defined by the analytic
continuation  of
\begin{equation}
\zeta(s)=\frac{1}{\Gamma(s)}\int_0^\infty \frac{t^{s-1}}{e^t-1}dt
\label{a6}
\end{equation}
for  Re$(s) <1$ \cite{elizalde94}.  As is known, $\zeta(s)$ is
one-valued everywhere except for $s=1$, where it has a single pole
with residue 1. As $\Gamma(s)$ has simple poles at $s=-n$ with
residue $(-1)^n/n!$, $n=0,1,2,...$ we get, by taking $c>4$ and
closing the contour of integral (\ref{a4}) as a large semicircle
on the left,
\begin{equation}
\beta F^{TE}=-C\int_0^\infty Bx\left[\frac{\Gamma(4)}{(2
ax)^4}\frac{1}{(2\pi C)^2}+\zeta(-1)-2 ax\sqrt{2\pi
C}\,\zeta(-\frac{3}{2})+...\right]dx. \label{a7}
\end{equation}
The first term in (\ref{a7}) diverges, the reason being that we
have replaced the lower limit $x_0$ by zero. This term
 is independent of $T$,  and can thus be
omitted in the present context.     Thus using $\zeta(-1)=-1/12,$
$\zeta(-3/2)=-0.025485$, together with the integrals
$\int_0^\infty Bxdx=1/12$ and $\int_0^\infty Bx^2dx=8/105$ gives
the temperature-dependent free energy to first order in $B$.

However, this is easily extended to arbitrary order in $B$ by
expansion of the integrand in (\ref{a1}) since $(B\,e^{-\alpha
x})^n= B^n(1-\alpha nx+\frac{1}{2}(\alpha nx)^2+\cdots)$. Thus by
summation $xB \rightarrow x\sum_{n=1}^\infty B^n/n=-x\ln(1-B)$ and
$x(Bx) \rightarrow x\sum_{n=1}^\infty (nB^n/n)x=x^2 B/(1-B)$. By
this the above integrals become those of Eqs.~(\ref{14}) and
(\ref{28}), and for the temperature dependent part of the free
energy we thus get
\begin{eqnarray}
\nonumber \Delta
F^{TE}=-\frac{C}{\beta}\left[-g'(0)\,\zeta(-1)-2a\sqrt{2\pi C} I\,
\zeta\left(-\frac{3}{2}\right)+\cdots\right]\\
=\frac{C}{\beta}\left[-\frac{1}{12}g'(0)\right]\left[1-8\,\zeta \left(-\frac{3}{2}\right)
\frac{3a\sqrt{2\pi C}}{-12 g'(0)}+\cdots\right].
\label{a11}
\end{eqnarray}
fully consistent with result (\ref{29}). Note that $\zeta(-3/2)$ is close to the approximate value (\ref{23}) and even closer to the more accurate value (\ref{24}).
Thus we have reason to consider $\zeta(-3/2)$ to be the exact result for the Euler-Maclaurin
expansion performed in Sec.~\ref{sec2}.

\setcounter{equation}{0}
\section{Remarks on the quantity $R$}

Let us explain in some more detail the interpretation of the
quantity $R$, defined in Eq.~(\ref{32}). This quantity gives the
relative difference between the temperature dependent theoretical
free energy $\Delta F_{th}^{TE}$, having the form (\ref{31}), and
the temperature dependent numerical free energy $\Delta
F_{num}^{TE}$ calculated from  Eq.~(\ref{6}). (As mentioned in
Sec.~3, the TM mode behaves as $T^4$ and is thus negligible near
$T=0$.)

Let us assume that $\Delta F_{th}^{TE}$ has the same form
(\ref{31}) as before, with coefficients $C_1$ and $C_2$, and that
$\Delta F_{num}^{TE}$ has the form
\begin{equation}
 \Delta F_{num}^{TE} = D_1 (T^2 - D_2 T^{5/2} +D_3 T^3+\cdots),
\end{equation}
with calculated values for the coefficients $D_1$, $D_2$, and $D_3$. Then,
\begin{equation}
 R= \frac{\Delta F_{th}^{TE}-\Delta F_{num}^{TE}}{\Delta F_{th}^{TE}}
 = \frac{C_1-D_1}{C_1} + \frac{D_1}{C_1}(D_2-C_2)T^{1/2} + \frac{D_1}{C_1}(C_2D_2-D_3) T +\cdots
\end{equation}
If $C_1=D_1$ and $C_2=D_2$, we see that $R$ is zero at $T=0$ and
is linear in $T$ for low $T$. From Fig.~3 we see that the fit is
perfect insofar as it may be determined from the graph. A constant
term would have caused a nonzero value at $T=0$, and a nonzero
$T^{1/2}$ term would have caused a vertical slope near $T=0$. None
of these effects are perceivable within the numerical accuracy,
from which we must conclude that $C_1$ and $C_2$ are correct within
numerical accuracy.

\end{document}